\documentclass[%
notitlepage,
twocolumn,
superscriptaddress,
nofootinbib,
amsmath,
amssymb,
aps,
pra,
]{revtex4-1}
\usepackage{amsfonts}
\usepackage{graphicx}
\usepackage{physics}
\usepackage{pgf}
\usepackage{bm}
\usepackage{soul}
\usepackage{color}
\usepackage{xcolor}
\usepackage{dsfont}
\usepackage{comment}
\usepackage{ulem}
\usepackage{hyperref}
\hypersetup{
	citecolor=blue,
	colorlinks=true,
	urlcolor=magenta,
}

\newcommand{\ICFO}{ICFO - Institut de Ciencies Fotoniques, The Barcelona Institute of Science and Technology, Av. Carl Friedrich Gauss 3, 08860 Castelldefels (Barcelona), Spain}

\newcommand{\IOPPAS}{Institute of Physics PAS, Aleja Lotnikow 32/46, 02-668 Warszawa, Poland}

\newcommand{\VILNIUS}{Institute of Theoretical Physics and Astronomy,
	Vilnius University, Saul\.etekio 3, LT-10257, Vilnius, Lithuania}

\newcommand{\OIST}{Quantum Systems Unit, Okinawa Institute of Science and Technology Graduate University, Onna, Okinawa 904-0495, Japan}

\begin{document}
	
\title{
Spin squeezing in open Heisenberg spin chains
}
	
\author{T. Hern\'andez Yanes}
\affiliation{\IOPPAS}
\author{G. \v{Z}labys}
\affiliation{\VILNIUS}
\affiliation{\OIST}
\author{M. P\l{}odzie\'n}
\affiliation{\ICFO}
\author{D. Burba}
\affiliation{\VILNIUS}
\author{M. Mackoit Sinkevi\v{c}ien\.e}
\affiliation{\VILNIUS}
\author{E. Witkowska}
\affiliation{\IOPPAS}
\author{G. Juzeli\=unas}
\affiliation{\VILNIUS}

\date{\today}
	
\begin{abstract}
Spin squeezing protocols successfully generate entangled many-body quantum states, the key pillars of the second quantum revolution. 
In our recent work [Phys. Rev. Lett. 129, 090403 (2022)] we showed that spin squeezing described by the one-axis twisting model can be generated in the Heisenberg spin-1/2 chain with periodic boundary conditions when accompanied by a position-dependent spin-flip coupling induced by a single laser field. 
%
%
In this work, we show analytically that the change of boundary conditions from the periodic to the open ones significantly modifies spin squeezing dynamics. A broad family of twisting models can be simulated by the system in the weak coupling regime, 
including
the one- and two-axis twisting under specific conditions, providing the Heisenberg level of squeezing and acceleration of the dynamics. 
Our analytical findings are confirmed by full numerical simulations.
\end{abstract}

\maketitle 

\section{Introduction}

Neutral atom arrays have recently emerged as promising platforms for realizing programmable quantum systems~\cite{Bernien2017, Browaeys2020, doi:10.1126/science.aal3837}. Based on individually trapped cold atoms in optical lattices~\cite{PhysRevLett.81.3108} and tweezers with strong
interactions between Rydberg states~\cite{Kaufman2021}, atom arrays
have been utilized to explore physics involving Hubbard and Heisenberg
models~\cite{RevModPhys.80.885, Joordens2008, Hart2015, Simon2011,PhysRevLett.115.215301}. It has been shown that indistinguishable Hubbard bosons serve as a platform for the generation and storage of metrologically useful many-body quantum states~\cite{Plodzien2020, https://doi.org/10.48550/arxiv.2208.04019, Plodzien2022,PRXQuantum.2.030329}. In some regime of parameters, arrays of ultra-cold atoms simulate chains of distinguishable spins (qubits) which are perfectly suitable for quantum information tasks and the generation of massive non-classical correlations, including 
Bell correlations and non-locality~\cite{Acin_2018,Eisert2020,Kinos2021,Laucht_2021}. These quantum many-body systems are crucial resources for emerging quantum technologies~\cite{Zwiller2022,Fraxanet2022}. 

\begin{figure}
	\centering\includegraphics[width=0.45\textwidth]{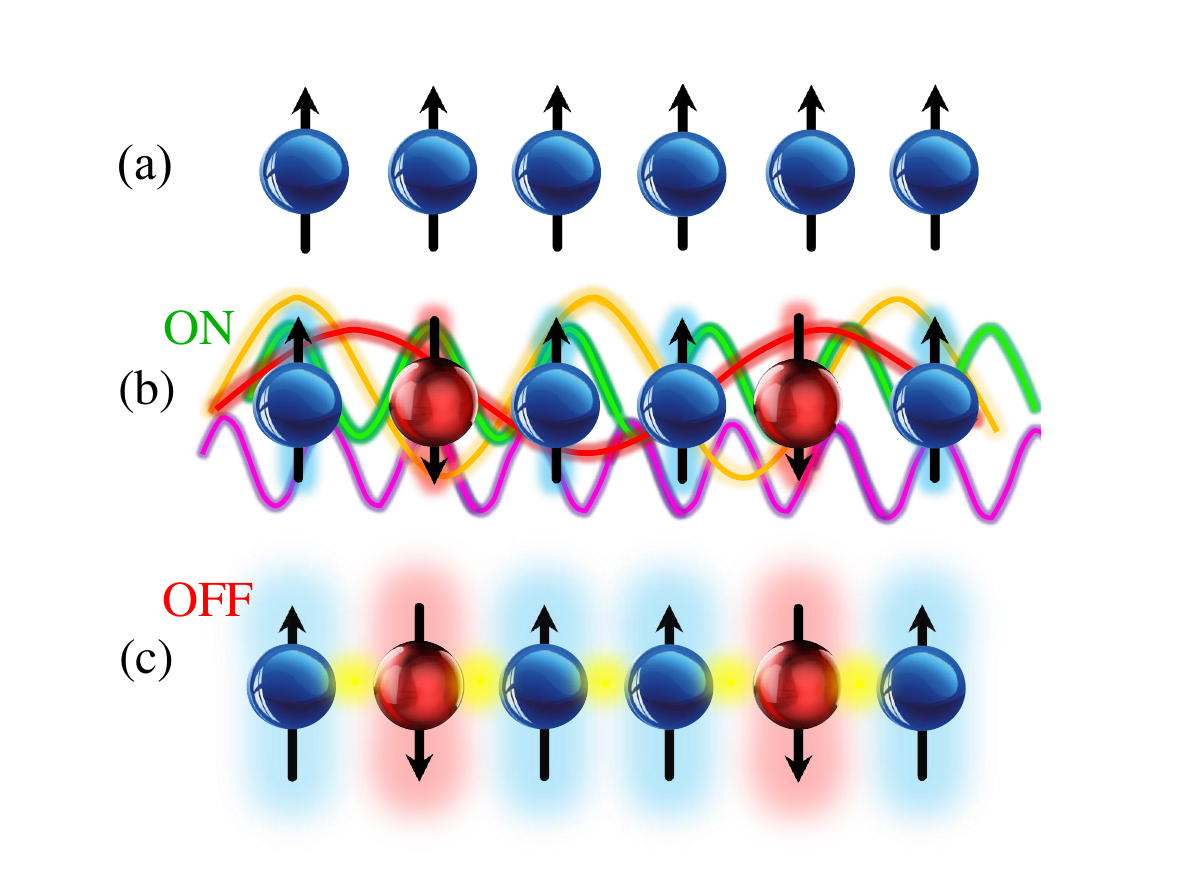}
	\caption{Illustration of the Ramsey-type spectroscopy scheme. (a) Preparation of the initial spin coherent state. (b) The excitation of spin waves states (different color lines) by the spin-flip coupling 
 serves as an intermediate state to induce "effective" interaction and establish correlations between elementary spins. (c) Turning off the coupling freezes the dynamics, and the spin-squeezed states are stored in the Mott insulating phase. Panels (b) and (c) illustrate an example of a configuration of spins. Yet, the resulting state during and at the end of evolution is a superposition of various possible configurations including the initial one presented in (a). }
	\label{fig:fig0}
\end{figure}

Systems composed of ultra-cold fermions in optical lattices have also attracted a lot of attention currently in the context of the generation of non-classical states, see e.g. in~\cite{PhysRevResearch.1.033075, PhysRevResearch.3.013178, PhysRevLett.129.113201}.
In particular, in our recent work \cite{PhysRevLett.129.090403}, we have shown that in a lattice of strongly interacting ultra-cold fermionic atoms involving two internal states, it is possible to generate non-classical correlations when adding position-dependent atom-light coupling. 
The Fermi-Hubbard model describing the system under periodic boundary conditions (PBC) can be cast onto an isotropic spin-1/2 Heisenberg chain in a deep Mott regime, while the atom-light coupling can be considered as a position-dependent spin-flipping.
To generate spin squeezing the Ramsey-type spectroscopy scheme is considered~\cite{PhysRevLett.129.090403}, as illustrated in Fig.~\ref{fig:fig0}. 
As soon as the atoms are put in a coherent superposition of two internal states by an electromagnetic pulse, an additional weak atom-laser coupling is turned on.
This coupling activates the general mechanism in PBC case: it induces excitation of a pair of spin waves with opposite quasi-momentum. These spin waves extend over the entire system allowing individual atoms to interact "effectively" and establish non-trivial quantum correlations~\cite{PhysRevLett.82.1835,PhysRevResearch.1.033075,PhysRevResearch.3.013178,PhysRevLett.126.160402, PhysRevLett.129.090403}. 
When the desired level of spin squeezing is established, the spin-flip coupling is turned off but the quantum correlations survive and are stored deeply in the Mott insulating phase.
We showed that the isotropic Heisenberg spin-1/2 chain with the weak position-dependent spin-flip coupling generates spin-squeezing dynamics given by the one-axis twisting (OAT) model. 
Furthermore, we numerically observed that open boundary conditions (OBC) change the spin squeezing dynamics. Depending on the coupling parameters, an acceleration of squeezing generation was observed with the same or similar level of squeezing~\cite{PhysRevLett.129.090403}. 

In this paper, we provide a detailed analytical and numerical analysis of the impact of OBC on the spin squeezing dynamics in Heisenberg spin chains. To this end, we develop the spin-waves theory for OBC by modifying the coordinate Bethe ansatz~\cite{Bethe1931}. Next, by using the Schrieffer-Wolf transformation~\cite{Chao_1977,PhysRevB.18.3453,PhysRev.149.491,BRAVYI20112793,PhysRevResearch.3.013178} we derive the effective model in terms of collective spin operators to describe the squeezing dynamics generated in the weak coupling regime.
For OBC the coupling leads to the excitation of a superposition of spin waves with different energies and amplitudes rather than a pair of spin waves with opposite quasi-momentum, as it is the case for PBC. This still allows individual atoms to correlate and generate squeezing. However, the excitation of a superposition of spin waves complicates the form of the effective model.  
We analyze this unconventional model in detail identifying the initial conditions and the coupling parameters for spin squeezing generation with the level given by the OAT and two-axis counter twisting (TACT) models~\cite{PhysRevA.47.5138, PhysRevA.92.013623, PhysRevLett.129.090403}. 
Consequently, we show that it is possible to generate Heisenberg level of squeezing in spin-1/2 Heisenberg chains under OBC. 
In addition, we show that the corresponding time scale of the best squeezing is reduced with respect to PBC when keeping the same perturbation level. 
Our analytical findings were confirmed by full numerical simulations.
The results obtained can be used in
the current state-of-the-art experiments with ultra-cold atoms in optical lattices~\cite{Campbell3987, Bromley_2018,Bataille_2020} and tweezer arrays~\cite{Young2020, https://doi.org/10.48550/arxiv.2110.15398}.


\section{Heisenberg model and spin-waves states for OBC}
\label{sec:model}

Let us concentrate on a specific physical system composed of the total even number $N$ of fermionic ultra-cold atoms loaded into a one-dimensional optical lattice potential of $N$ sites. Each atom has two internal states $\ket{\uparrow}$ and $\ket{\downarrow}$ corresponding to a spin-1/2 degree of freedom. 
The atoms are assumed to occupy the lowest Bloch band, interact through~s-wave collisions, and hence can be described by the Fermi-Hubbard model. 

We assume the interaction dominates over the tunnelling and the system is in the Mott insulating phase at half-filling when double occupancy of a single site is energetically unfavourable. 
The second order processes, obtained by a projection onto the manifold of single occupancy of lattice sites, lead to the nearest-neighbour spin-exchange interactions~\cite{PhysRevLett.129.090403,PhysRevResearch.3.013178,Chao_1977,PhysRevB.18.3453,PhysRev.149.491,BRAVYI20112793,PhysRevResearch.3.013178}. The spin dynamics of this system is well captured by the isotropic Heisenberg (spin exchange) model~\cite{Heisenberg1928, Demler2003}
\begin{equation}\label{eq:HSE}
\hat{H}_{\rm SE} = J_\text{SE} \sum_{j=1}^{N-1} 
\left(
\hat{S}^x_{j} \hat{S}^x_{j+1} + \hat{S}^y_{j} \hat{S}^y_{j+1} + \hat{S}^z_{j} \hat{S}^z_{j+1} - \frac{1}{4}
\right),
\end{equation}
where $J_{\rm SE}$ represents the spin-exchange energy, 
$\hat{S}^{+}_j = \hat{a}^\dagger_{j,\uparrow}\hat{a}_{j,\downarrow}$, $\hat{S}^{-}_j =\hat{a}^\dagger_{j,\downarrow}\hat{a}_{j,\uparrow}$, $\hat{S}^{\pm}_j = \hat{S}^{x}_j \pm i\hat{S}^y_j$, $\hat{S}^{z}_j = (\hat{n}_{j,\uparrow}-\hat{n}_{j,\downarrow})/2$ are on-site spin operators, and where we take $\hbar=1$.
The fermionic operators $\hat{a}_{j, s}$ annihilate an atom in the $j$th lattice site in the state $s\in \{\uparrow,\downarrow\}$, and  $\hat{n}_{j,s}=\hat{a}^\dagger_{j, s}\hat{a}_{j,s}$ is the corresponding on-site operator of the number of atoms. We also introduce the collective spin operators $\hat{S}_\sigma = \sum_j \hat{S}_j^\sigma$ with $\sigma=x,y,z,\pm$. 
The analytical form of the energy spectrum of the Hamiltonian~(\ref{eq:HSE}) and corresponding eigenstates for PBC are known from 1931 due to the famous work of Bethe~\cite{Bethe1931}. Their counterpart for OBC is less explored, up to our knowledge. 

The Hamiltonian \eqref{eq:HSE} is spherically symmetric with respect to spin rotation. Thus eigenstates of $\hat{H}_{\rm SE}$ can be taken to be also the eigenstates of the square of the total spin  $\hat{S}^2 = \hat{S}_x^2+\hat{S}_y^2+\hat{S}_z^2$ and its $z$ projection $\hat{S}_z$ 
with the eigenvalues $S(S+1)$ and $m$, respectively. To understand the spin squeezing dynamics let us first recall the analytical form of two energy manifolds of $\hat{H}_{\rm SE}$ characterized by the largest values of the total spin.

The first energy manifold corresponding to the total spin quantum number $S=N/2$ is spanned by Dicke states $|m\rangle \equiv |N/2, m\rangle$ which are zero energy eigenstates of $\hat{H}_{SE}$. They can be represented in terms of the  all spins up state affected $N/2-m$ times by the collective spin lowering operator $\hat{S}_-$: 
\begin{equation}
\label{Dicke-states}
|{m}\rangle=\sqrt{\frac{(N/2+m)!}{(N/2-m)!(N )!}} \hat{S}_-^{N/2-m} \bigotimes_{j=1}^N \ket{\uparrow}_j,
\end{equation}
where the quantization axis is chosen to be along the $z$ direction: $\hat{S}_j^z \ket{\uparrow}_j= 1/2 \ket{\uparrow}_j$ and $\hat{S}_j^z \ket{\downarrow}_j= -1/2 \ket{\downarrow}_j$. 
Alternatively, the Dicke states $|{m}\rangle$ can be defined by using the rising operator $\hat{S}_+\equiv(\hat{S}_-)^\dagger$ in the place of $\hat{S}_-$ when replacing $m$ and $\ket{\uparrow}_j$ with $-m$ and $\ket{\downarrow}_j$, respectively, on the right-hand side of (\ref{Dicke-states}).
The  Dicke states are eigenstates of  $\hat{H}_{\rm SE}$ with zero eigen-energies  for both PBC and OBC. Altogether there are $N+1$ Dicke states corresponding to different values of $m\in (-N/2, -N/2 +1,\cdots, N/2)$. 

The second energy manifold to be considered is spanned by the spin-wave states~\cite{Lamers_2015,gaudin_2014,PhysRevLett.129.090403,PhysRevResearch.3.013178} containing one spin excitation and characterized by the total spin quantum number $S=N/2 - 1$. In the case of OBC one can solve analytically the eigenproblem of these states for the Hamiltonian (\ref{eq:HSE}) by using the coordinate Bethe ansatz modified appropriately to account for the difference coming from the two boundary points, see Appendix~\ref{app:sws} for derivation. This leads to the following form of the spin-wave states
\begin{equation}\label{eq:swsobc}
|m, {q} \rangle = \pm \sqrt{N} c_{N/2, \pm m} \sum_{j=1}^N p^{(q)}_j \hat{S}^{\pm}_j | m\mp 1\rangle,
\end{equation}
where
\begin{equation}\label{eq:cn2mpm1}
c_{N/2, \pm m}=\sqrt{\frac{N-1}{(N/2\mp m)(N/2\mp m+1)}}. 
\end{equation}
The sign $\pm$ in Eq.~\eqref{eq:swsobc} for $|m, {q} \rangle$ corresponds to two equivalent definitions of the spin waves in terms of the on-site spin raising and lowering operators $\hat{S}_j^\pm$ acting on the Dicke states. Furthermore, the coefficients featured in Eq.~\eqref{eq:swsobc} are
\begin{equation}\label{eq:p_j}
p^{(q)}_j = \sqrt{\frac{2}{N}} \cos\left[\frac{\pi}{N}\left( j -\frac{1}{2}\right)q \right]\,.
\end{equation}
Altogether there are $(N-1)^2$ different spin-wave states corresponding to various combinations of quantum numbers $m\in (-N/2+1, -N/2 +2,\cdots, N/2-1)$ and $q=1,2,\cdots, N-1$.
The corresponding eigenenergies $E_{q}$  do not depend on the spin projection quantum number $m$ and read
\begin{equation} \label{eq:E_q}
E_{q} = J_{SE} \left[\cos (\frac{\pi}{N} q) - 1\right].
\end{equation}
Notice, that for OBC the amplitudes $p^{(q)}_j$ given by Eq.~\eqref{eq:p_j} represent standing waves. They thus differ from the solution for PBC where the amplitudes $p^{(q)}_j=N^{-1/2} e^{i 2\pi q j/N} $ are plane waves~\cite{gaudin_2014}. This has substantial consequences for the coupling mechanism and the spin squeezing dynamics analyzed in Sections~\ref{sec:effectivehamiltonian} and \ref{sec:spinsqueezingforOBC}.

\section{Protocol for dynamical generation of spin squeezing}
\label{sec:protocol}

In order to generate spin squeezing in this Heisenberg spin-1/2 chain with OBC described by Hamiltonian~(\ref{eq:HSE}) we add an atom-light coupling which induces position-dependent spin-flipping. The resulting system Hamiltonian $\hat{H}_{\rm spin}$ reads
\begin{align}\label{eq:Heff}
\hat{H}_{\rm spin} & = \hat{H}_{\rm SE} +  \hat{H}_{\uparrow\downarrow}, \\
\hat{H}_{\uparrow\downarrow} & =  \frac{\Omega}{2} \sum_{j=1}^N
\left(e^{i (\phi j - \phi_0)} \hat{S}^+_{j} +  e^{-i (\phi j - \phi_0)} \hat{S}^-_{j}\right)\,,\label{eq:SFCj}
\end{align}
where the extra term $\hat{H}_{\rm \uparrow \downarrow}$ represents the sum over the on-site spin-flip coupling with the amplitude $\Omega$ and position-dependent phase~$\phi j$, where $\phi=\pi \cos({\alpha}) \lambda_{\mathrm{latt}}/\lambda_L $ can be tuned by properly choosing an angle $\alpha$ between laser beams producing the optical lattice and the direction of laser field inducing the coupling. The two beams are characterized by 
the wave-lengths $\lambda_{\mathrm{latt}}$ and $\lambda_L$, respectively, see e.g. in~\cite{PhysRevLett.129.090403}. 
Here, $\phi_0\in[0, 2 \pi)$ is the global off-set phase of the coupling lasers, which can be interpreted as the transformation of $\hat{H}_{\uparrow\downarrow}$ due to the global spin rotation around the $z$ axis by the angle $\phi_0$. Equivalently, it can also be interpreted as the spin rotation for the initial state around the same $z$ axis and by the same angle $\phi_0$, but in the opposite direction. 

In the case of PBC, the coupling phase $\phi$ should be commensurate with $2 \pi /N$, namely $\phi=2 \pi n/N$, where $n=1,2,\cdots, N-1$, to ensure periodicity of $\hat{H}_{\uparrow \downarrow}$~\cite{PhysRevLett.129.090403}. 
Here, however, we are interested in OBC, and therefore $\phi$ can take any real values apart from the trivial one $\phi =0$ or $\phi = 2\pi$ for which $\hat{H}_{\rm \uparrow \downarrow}$ does not provide coupling between the Dicke and the spin-wave state manifolds needed for the generation of spin squeezing. 

The initial state convenient to start the evolution is the spin coherent state
\begin{equation}\label{eq:scs}
|\theta, \varphi \rangle = e^{-i \hat{S}_z\varphi} e^{-i \hat{S}_y \theta} \bigotimes_{j=1}^N \ket{\uparrow}_j  ,
\end{equation}
where all the spins point in the same direction parameterized by the spherical angles $\theta$ and $\varphi$.
In general, the spin-coherent state (\ref{eq:scs}) belongs to the Dicke manifold of the total spin $S=N/2$ and hence can be expressed in the basis of the Dicke states \eqref{Dicke-states} as 
\begin{equation}
    |\theta, \varphi \rangle = \sum_{m=-N/2}^{N/2} a_m| m \rangle,
\end{equation}
where 
\begin{equation}
    a_m=\sqrt{\binom{N}{\frac{N}{2}+m}} \cos^{\frac{N}{2}+m}\left(\frac{\theta}{2}\right) \sin^{\frac{N}{2}-m}\left(\frac{\theta}{2}\right) e^{i(\frac{N}{2}-m)\varphi}
\end{equation}
are coefficients of decomposition.

The subsequent evolution of the initial state is defined by the unitary operator $\hat{U} = e^{-i t \hat{H}_{\rm spin}}$. To quantify the level of squeezing generated in time we use the spin squeezing parameter
\begin{equation}\label{eq:ssqparameter}
\xi^2 = \frac{N (\Delta \hat{S}_{\perp})_{\rm  min}^2}{\langle \hat{S} \rangle^2}
\end{equation}
where the length of the mean collective spin is $\langle \hat{S} \rangle$ and the minimal variance of the collective spin orthogonally to its direction is $(\Delta \hat{S}_{\perp})_{\rm  min}^2$~\cite{PhysRevA.46.R6797}. 

Non-trivial quantum correlations are produced in the weak coupling regime, where the characteristic energy of the coupling Hamiltonian $\hat{H}_{\uparrow \downarrow}$ is smaller than that of the spin-exchange term $\hat{H}_{\rm SE}$. In the next section, we derive the effective model describing the spin squeezing dynamics in terms of collective spin operators.

\section{Effective model}
\label{sec:effectivehamiltonian} 

When the spin-flip coupling is weak compared to the energy of the spin exchange, the dynamics of the initial spin coherent state $|\theta, \varphi \rangle$ governed by the spin Hamiltonian $\hat{H}_{\rm spin}$ within the Dicke manifold can be well approximated using perturbation theory. Therefore, the coupling term $\hat{H}_{\uparrow \downarrow}$ can be treated as a perturbation. 
For reasons that will be explained later, let us 
rephrase this operator in the following way:
\begin{equation}\label{eq:generalSOC-1}
\hat{H}_{\uparrow \downarrow} = 
\hat{\tilde{H}}_{\uparrow \downarrow}
+
v_x \hat{S}_x 
+
v_y \hat{S}_y,
\end{equation}
where
\begin{equation}\label{eq:generalSOC-1'}
\hat{\tilde{H}}_{\uparrow \downarrow} = 
\frac{\Omega}{2}
\sum_{j=1}^N
\left( 
\alpha^{+}_j \hat{S}^+_{j}
+ 
\alpha^{-}_j\hat{S}^-_{j} 
\right)\,.
\end{equation}
Here, $\alpha^{\pm}_j = e^{\pm i (\phi j - \phi_0)}-A^\pm$ with $A^\pm=\frac{1}{N} \sum_j e^{\pm i (\phi j - \phi_0)}$, as well as $v_x = \Omega {\rm Re}[A^+]/2 $ and $v_y = -\Omega{\rm Im}[A^+]/2 $. The separation of the two last terms in (\ref{eq:generalSOC-1}) is made in such a way that $\alpha^{\pm}_j$ sum up to zero.
Notice, $v_x$ and $v_y$ are non-zero only for phases~$\phi$ incommensurate with $2\pi/N$.

\subsection{First and second order contributions}

The operator $\hat{\tilde{H}}_{\uparrow\downarrow}$ on the right-hand side of (\ref{eq:generalSOC-1}) induces the coupling between the Dicke and spin-wave state manifolds while the remaining ones directly couple the Dicke states and represent the first-order perturbation term
\begin{equation}
\label{eq:H_eff^1}
\hat{H}_{\rm eff}^{(1)} = v_x \hat{S}_x + v_y \hat{S}_y\,. 
\end{equation}
To generate spin squeezing one needs to take into account the second-order contribution induced by~$\hat{\tilde{H}}_{\uparrow \downarrow}$. It can be obtained via the Schrieffer-Wolf transformation~\cite{PhysRevLett.129.090403,PhysRevResearch.3.013178,Chao_1977,PhysRevB.18.3453,PhysRev.149.491,BRAVYI20112793,PhysRevResearch.3.013178} leading to
\begin{equation}\label{eq:HSWgeneral}
    \hat{H}_{\rm eff}^{(2)} = \hat{I}_{N/2} \hat{\tilde{H}}^{\uparrow\downarrow} \hat{G}_{N/2-1} \hat{\tilde{H}}^{\uparrow\downarrow}
    \hat{I}_{N/2} ,
\end{equation}
where $\hat{I}_{N/2}=\sum_m |m \rangle \langle m|$ is the unit operator for projection onto the Dicke manifold, while $\hat{G}_{N/2-1}=\sum_{q\ne 0, m} \frac{|m,q\rangle\langle m, q|}{-E_{q}}$ is an operator which sums projectors onto the spin-wave states manifold with the corresponding energy mismatch denominator $- E_{q}$.
The matrix elements of (\ref{eq:HSWgeneral}) are
\begin{equation}\label{eq:matrixheff}
\langle m'| \hat{H}^{(2)}_{\rm eff} |  m\rangle = 
- \sum_{m'',q} 
\frac{\langle { m'}|\hat{\tilde{H}}_{\uparrow \downarrow}|m'', q \rangle
	\langle m'', q |\hat{\tilde{H}}_{\uparrow \downarrow}| m\rangle}{E_{q}}.
\end{equation}
Details about the transformation and its application to the Heisenberg spin-1/2 chain with the spin-flip coupling can be found in the Supplementary Material of reference~\cite{PhysRevLett.129.090403}. 
In the following, we focus on the derivation of the effective Hamiltonian $\hat{H}^{(2)}_{\rm eff}$ and its representation in terms of the collective spin operators.

\begin{figure}
\centering
	\includegraphics[width=\columnwidth]{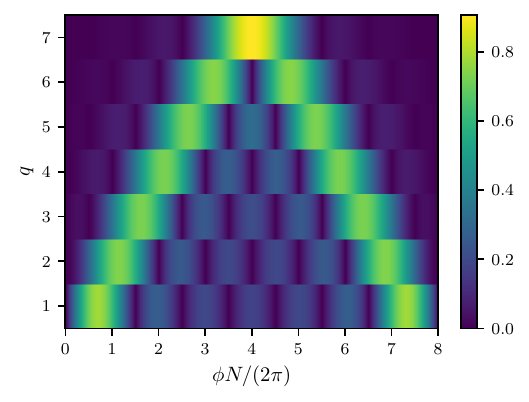}
	\caption{
	The absolute values of the normalized coefficients $|f^{+}_q|{N}^{-1/2}$ are shown by color versus the coupling phase $\phi\in \mathbb{R}$ and the spin-waves quantum number $q\in\mathbb{Z}$ for an arbitrary $\phi_0$ when $N=8$. 
	}
	\label{fig:fig2}
\end{figure}

Let us start with expressing the action of $\hat{\tilde{H}}_{\uparrow \downarrow}$ on Dicke states, namely
\begin{align}
\hat{\tilde{H}}_{\uparrow \downarrow}| m\rangle 
&= \frac{\Omega}{2}
|\Psi, m+1\rangle^+ +
\frac{\Omega}{2}
|\Psi, m-1\rangle^-\,,
\label{eq:swscoupling}
\end{align}
where states $|\Psi, m \pm 1\rangle^\pm =\sum_j \alpha^{\pm}_j \hat{S}^\pm_j |m\rangle $ can be expanded in terms of the spin-wave states $|m\pm1,q \rangle$ as
\begin{align}
\label{eq:psiP}
|\Psi, m\pm 1\rangle^\pm &=
\sqrt{N}c_{N/2, \pm m+1} \sum_q f^{\pm}_{q} |m\pm 1,q \rangle.
\end{align}
Here, $c_{N/2, m\pm 1}$ are given by Eq.~(\ref{eq:cn2mpm1}) and
\begin{equation}
\label{eq:f_q}
f^{\pm}_{q} =  \sum_j p^{(q)}_j \alpha^{\pm}_j = \sum_j p^{(q)}_j e^{\pm i (\phi j - \phi_0)}\,,
\end{equation}
with $f^{+}_{q}=(f^{-}_{q})^*$ because $p^{(q)}_j$ is real. Note, the spin-flip term $\hat{\tilde{H}}_{\uparrow \downarrow}$ couples each Dicke state $|m\rangle$ with a superposition of spin-wave states (\ref{eq:psiP}) characterized by energies $E_q$. This is different from the PBC case where $\hat{H}_{\uparrow \downarrow}$ couples each Dicke state with a pair of spin-wave states of well-defined quantum numbers $q=\pm \phi N/(2 \pi)$ set by the coupling phase $\phi$~\cite{PhysRevLett.129.090403}. An example of the amplitude of elementary couplings $f^{+}_q$ to the $|m,q\rangle$ states is presented in Fig.~\ref{fig:fig2}.
We can see that, indeed, the coupling could be non-negligible even to the lowest state $|m,q=1\rangle$. Therefore, the perturbative regime is defined by
the smallest energy gap, namely $\Omega \ll |E_{q=1}|=J_{\rm SE}|\cos(\pi/N)-1|$.

The relevant matrix elements of the second-order  contribution can be written as
\begin{align}
\langle m'', q|\hat{\tilde{H}}_{\uparrow\downarrow} |m\rangle 
&= \frac{\Omega}{2} N^{-1/2} c_{N/2, m+1}^{-1} f^{+}_{q}
\delta_{m'', m+1} \nonumber \\
&+
\frac{\Omega}{2} N^{-1/2} c_{N/2, -m+1}^{-1} f^{-}_{q}
\delta_{m'', m-1},
\label{eq:matrixel}
\end{align}
where the coefficients $N^{-1/2}c_{N/2, \pm m+1}^{-1}$ come from the scalar product between the Dicke state $|m\rangle$ and the states $|\Psi, m\pm 1\rangle^\pm$. The non-zero matrix elements of the second-order term~(\ref{eq:matrixheff}), namely $H_{m',m}=\langle m'| \hat{H}^{(2)}_{\rm eff} | m\rangle$, read
\begin{align}
	H_{m, m}&= 
	- ( c_{N/2, m}^{-2} +
	c_{N/2, -m}^{-2}) \, (N-1) \chi_z ,\label{eq:effdiag}\\
	H_{m, m-2}&= c_{N/2, m-1}^{-1} c_{N/2, -(m-1)}^{-1} \, (N-1) \chi_x, 
    \label{eq:effoff-2}\\
	H_{m, m+2}&= c_{N/2, m+1}^{-1} c_{N/2, -(m+1)}^{-1}\, (N-1) \chi_x,
    \label{eq:effoff+2}
\end{align}
where 
\begin{align}
    \chi_z = \frac{\Omega^2}{4 N J_{\rm SE}(N-1)} \sum_{q=1}^{N-1}  \frac{f^{+}_{q} f^{-}_{q}}{\cos (\frac{\pi}{N} q) - 1} , \label{eq:Fdiag}\\
    \chi_x = \frac{\Omega^2}{4 N J_{\rm SE}(N-1)} \sum_{q=1}^{N-1} \frac{\left(f^{-}_{q} \right)^2}{\cos (\frac{\pi}{N} q) - 1}.
    \label{eq:FdiadFodd}
\end{align}
Comparing the matrix elements presented in Eqs.~(\ref{eq:effdiag})-(\ref{eq:effoff+2}) with the matrix elements of the appropriate collective spin operators, the second-order perturbation contribution can be represented in the operator form as
\begin{align}
    \hat{H}^{(2)} _{\rm eff}&= 
    - 2 \chi_z \left(
    \hat{S}^2 + \hat{S}_z^2\right) 
    + {\rm Re} \left[\chi_x\right]
    \left( \hat{S}^2_+ + \hat{S}^2_-\right)\nonumber \\
    & + i {\rm Im} \left[\chi_x\right] \left( \hat{S}^2_+ - \hat{S}^2_-\right) ,
    \label{eq:fullsecond}
\end{align}
as explained in Appendix~\ref{app:matrixspins}.
The full effective Hamiltonian is a sum of the first- and second-order contributions:
\begin{equation}
\hat{H}^{(\phi_0)}_{\rm eff} = \hat{H}^{(1)}_{\rm eff} + \hat{H}^{(2)}_{\rm eff}.\label{eq:full28}
\end{equation}

\subsection{Choosing the off-set phase $\phi_0=\phi(N+1)/2$}

In what follows, we will take a value of the global coupling phase to be $\phi_0=\phi(N+1)/2$, so that $v_y$ entering Eqs.~\eqref{eq:generalSOC-1} and (\ref{eq:H_eff^1}), as well as the imaginary part of $\chi_x$ vanish, i.e.~
$v_y={\rm Im} \left[\chi_x\right]=0$, see Appendix~\ref{app:phi0}. This simplifies the form of the effective model leading to
\begin{equation}
\hat{H}^{(\phi`_0)}_{\rm eff} = 
-2\chi_z \left(\hat{S}^2 + \hat{S}_z^2 
- \eta \hat{S}^2_x
+ \eta \hat{S}^2_y +
\gamma \hat{S}_x\right),
\label{eq:effective-fin}
\end{equation}
where $\eta = \chi_x/\chi_z$ and $\gamma =v_x/\chi_z$. This specific choice of phase $\phi_0$ does not involve a loss of generality as the full effective Hamiltonian (\ref{eq:full28}) containing $\hat{H}^{(1)}_{\rm eff}$ and $\hat{H}^{(2)}_{\rm eff}$ of Eqs.~(\ref{eq:H_eff^1}) and (\ref{eq:fullsecond}) is related to that given by Eq.~(\ref{eq:effective-fin}) via a unitary transformation set by the global rotation around the $z$ axis through the angle $\phi_0$.

\begin{figure}
	\centering\includegraphics{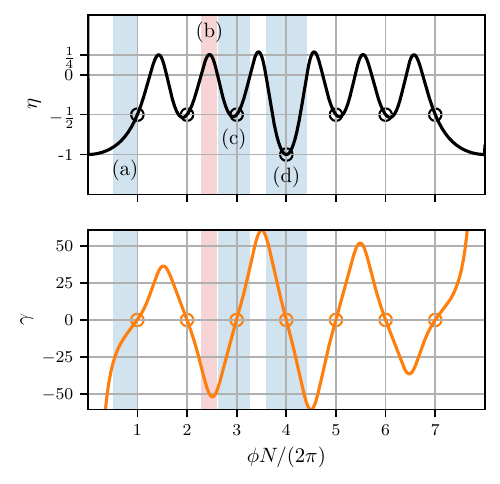}
	\caption{
	The parameters $\eta$ (top panel) and $\gamma$ (bottom panel) of the effective model (\ref{eq:effective-fin}) versus the coupling phase $\phi$ are marked by black and orange lines, respectively, for $N=8$, $\Omega = |E_{q=1}|/10$ and $\phi_0=\phi(N+1)/2$. The values of $\eta$ and $\gamma$ for commensurate phases are marked by open circles. The regions shaded in blue present examples when $\eta < 0$ while the one shaded in red when $\eta > 0$. 
 }
	\label{fig:fig3}
\end{figure}

In Fig.~\ref{fig:fig3} we show variation of the two parameters of the effective model~\eqref{eq:effective-fin}, namely $\eta$ and~$\gamma$, versus $\phi$. The commensurate phases corresponding to $\phi=2 \pi n/N$ with $n\in [1, N-1]$ are marked by open points in Fig.~\ref{fig:fig3} for which one has $\gamma=0$. 
In this case, we numerically observe that $\eta = - 1/2$ for $\phi \ne \pi$, and $\eta=-1$ for $\phi=\pi$. In addition, we have also analytically found that
\begin{align}
    \chi_z = -\frac{\Omega^2}{4 J_{\rm SE}(N-1)} \frac{2}{\cos (\phi) - 1} , \label{eq:Fdiag-comm}\\
    \chi_x = \frac{\Omega^2}{4 J_{\rm SE}(N-1)} \frac{1}{\cos (\phi) - 1},
    \label{eq:FdiadFodd-comm}
\end{align}
for commensurate phases $\phi = 2 \pi n/N$ apart from $\phi= \pi$ where 
\begin{equation}
    \chi_z = - \chi_x  = -\frac{\Omega^2}{4 J_{\rm SE}(N-1)}.\label{eq:chiz-pi}
\end{equation}
The derivation is presented in Appendix~\ref{app:triple-cos-sum}.
The non-commensurate coupling phases $\phi$ result in both positive and negative values of the parameter $\eta$ which is independent of $J_\mathrm{SE}$, $\Omega$, and $N$. On the contrary, the coefficient $\gamma$ depends on the system parameters, and scales as $\gamma \propto N J_\mathrm{SE} / \Omega$.

In this way, we derived the second-order contribution~(\ref{eq:fullsecond}), and consequently, the effective model~\eqref{eq:effective-fin} showing that the boundaries significantly modify the spin squeezing Hamiltonian with respect to PBC in which one arrives at the effective Hamiltonian in a form of the OAT model, namely $\hat{H}_{\rm eff} =-\chi_\pi \hat{S}_x^2$ for $\phi = \pi$ and $\hat{H}_{\rm eff} =\chi_\phi \hat{S}_z^2$ for $\phi\ne \pi$~\cite{PhysRevLett.129.090403}. Therefore, it is not only the time scale that is changed due to OBC but the entire dynamics as well. This is a counter-intuitive result as usually, the PBC describes well the system in the limit of large~$N$.

\section{Spin squeezing for OBC}
\label{sec:spinsqueezingforOBC}
 
In these subsections, we analyze the unitary evolution of spin squeezing parameter governed by the effective spin Hamiltonian~(\ref{eq:effective-fin}). We distinguish two cases depending on the commensurability of the coupling phase~$\phi$. 
We demonstrate that if the coupling phase is commensurate, the resulting model (\ref{eq:effective-fin}) can be either OAT for $\phi= \pi$ or non-isotropic TACT for $\phi\ne\pi$. The most general case of non-commensurate phases gives rise to the squeezing dynamics, however, not simulated by the conventional OAT and TACT twisting models.

\subsection{Spin squeezing with commensurate phase}
\label{subsec:commensurate}

Tuning the value of the coupling phase $\phi$ to the integer multiple of $2 \pi/N$ simplifies the problem. In particular, by taking $\phi = \pi$ we have $\eta=-1$ and the effective Hamiltonian (\ref{eq:effective-fin}) acquires the form of the OAT one, namely
\begin{equation}
\hat{H}_{\rm eff} = 
4  \chi_z \hat{S}^2_y,
\end{equation}
where we omitted a term proportional to~$\hat{S}^2$, as it only shifts the origin of energy. The convenient initial spin coherent states are the ones polarized in the $x-z$ plane, namely $|\theta, \varphi =0\rangle$ and for any $\theta$. The best level of squeezing~$\xi^2_{\rm best}\approx N^{-2/3}$ is achievable for times $t_{\rm best}\approx N^{-2/3}|4 \chi_z|^{-1}$, in the large $N$ limit according to the OAT dynamics~\cite{PhysRevA.47.5138,https://doi.org/10.48550/arxiv.2112.01786}. 
Next, taking the analytical expression (\ref{eq:chiz-pi}) for $\chi_z$ we obtain $t_{\rm best} \approx N^{1/3}J_{\rm SE}/\Omega^{2} $. Therefore, the twisting dynamics is essentially the same as for PBC~\cite{PhysRevLett.129.090403}. The only difference is that for OBC the resulting time scale is four times shorter compared to the PBC case when keeping the same perturbation level~$\Omega$. 
Acceleration of the best squeezing time takes place because of a broader range of amplitudes $p_j^{(q)}$ contributing to the generation of spin squeezing.

In another situation, when the coupling phase is not equal to $\pi$ we have $\eta=-1/2$ and $\gamma=0$, so the effective Hamiltonian (\ref{eq:effective-fin})  reduces to
\begin{equation}
\label{eq:effnepi}
\hat{H}_{\rm eff} = 
2 \chi_z \left( \hat{S}^2_y - \hat{S}^2_z/2 \right),
\end{equation}
where we omitted the term proportional to $\hat{S}^2$. Equation \eqref{eq:effnepi} represents the anisotropic TACT with the anisotropy equal to $1/2$. 
It is worth stressing here, that OBC provides anisotropic TACT without adding an extra atom-light coupling characterized by two different phases. In the case of PBC it was necessary to include two spin-flipping terms in order to simulate TACT~\cite{PhysRevLett.129.090403}. Let us consider again the initial state for the spin squeezing generation to be the spin coherent state polarized in the $x-z$ plane, $|\theta, \varphi =0\rangle$. The anisotropic TACT given by (\ref{eq:effnepi}) generates the Heisenberg limited level of squeezing $\xi^2_{\rm best}\approx N^{-1}$ on the time scale $t_{\rm best } \approx (2 \chi_z N\sqrt{2})^{-1} \ln (N/2)$~\cite{https://doi.org/10.48550/arxiv.2208.04019}. Therefore, taking into account the system parameters and the relation for $\chi_z$ given by~(\ref{eq:Fdiag-comm}) we have $t_{\rm best} \approx J_{\rm SE}{\rm ln} (N/2)|\cos\phi-1| /(\sqrt{2} \Omega^2 )$ which weakly depends on the system size $N$. 

\begin{figure}
    \centering
    \includegraphics[width=\columnwidth]{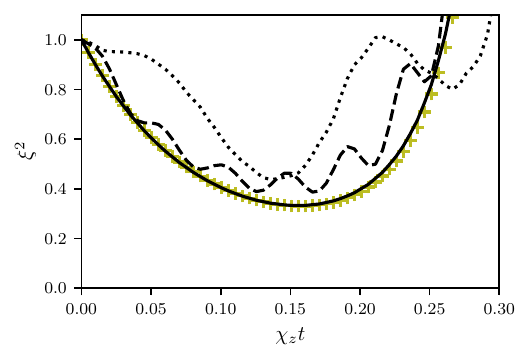}
    \caption{
    Variation of spin squeezing parameter (\ref{eq:ssqparameter}) in time for different values of $\Omega$ when the initial state is $|\theta=\pi/2,\varphi=0 \rangle $, $N = 8$ and $\phi = \pi - 2\pi/N,\, \phi_0 = \phi(N+1)/2$. The result for the effective model (\ref{eq:effective-fin}) is marked by olive crosses while results for the coupled Heisenberg model (\ref{eq:Heff}) are shown with black lines for $\Omega = |E_{q=1}|/10$ (solid), $\Omega = |E_{q=1}|$ (dashed) and $\Omega = 2 |E_{q=1}|$ (dotted).
    }
    \label{fig:fig4}
\end{figure}

\begin{figure}
	\centering
    \includegraphics[width=\columnwidth]{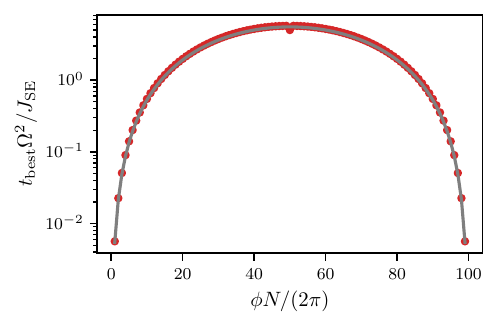}
	\caption{
	The best squeezing time $t_\mathrm{best}$ multiplied by $\Omega^2/J_\mathrm{SE}$ for $N=100$ to isolate dependence on the coupling phase $\phi$. The numerically evaluated values of the best squeezing time using unitary evolution according to (\ref{eq:effective-fin}) are shown with red points. The corresponding behaviour $t_{\rm best} \Omega^2/J_{\rm SE}={\rm ln} (N/2)|\cos\phi-1| /(\sqrt{2} \Omega^2 )$ for $\phi\ne \pi$ and $\phi_0 = \phi(N+1)/2$ is shown with a solid grey line, see text for more details.
    }
	\label{fig:fig5}
\end{figure}
\begin{figure*}
\centering
    \includegraphics{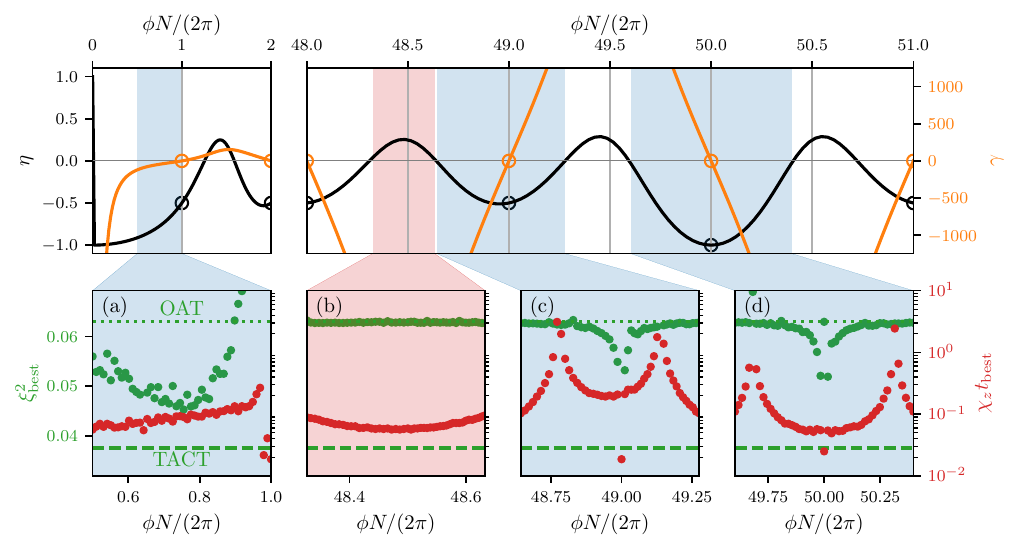}
    \caption{
    The best squeezing $\xi^2_\mathrm{best}$ (green points) and the best squeezing time $t_\mathrm{best}$ (red points) are shown in panels (a)-(d) for different regions of $\phi$. The numerical results for the effective model (\ref{eq:effective-fin}) with $N=100, J_\mathrm{SE} = 1$, $\Omega = |E_{q=1}|/10$, $\phi_0 = \phi(N+1)/2$ and $\eta > 0$ as indicated by the red shadowing areas and $\eta < 0$ indicated by the blue ones. The numerical values of $\eta$ and $\gamma$ used in simulations are shown in the top panels. The two limit cases for the values of $\xi^2_\mathrm{best}$, namely OAT and TACT for $N=100$, are marked with horizontal green dotted dashed lines, respectively.
    }
    \label{fig:fig6}
\end{figure*}

In Fig.~\ref{fig:fig4} we show examples of spin squeezing dynamics for different values of $\Omega$. A perfect agreement with the effective model (29) is observed in the perturbative regime when $\Omega \ll |E_{q=1}|$. Significant spin squeezing can also be generated beyond this regime, yet large discrepancies arise with respect to the TACT dynamics.

It is also worth commenting here on the importance of the coupling phase $\phi$ on the best squeezing time. The dependence on $\phi$ is hidden in the function~$\chi_z$. In Fig.~\ref{fig:fig5} we plotted the variation of the best squeezing time with the phase $\phi$. We can see the time scale increases by orders of magnitude for values of the coupling phase from $\phi = 2 \pi/N$ to $\phi = \pi$ and then decreases symmetrically to $\phi = 2\pi(N-1)/N$. Thus, in practical applications, the optimization of the system parameters $J_{\rm SE},\, \Omega,\, \phi$ will be necessary to have the shortest possible time scale.

\subsection{Spin squeezing with non-commensurate phases}
\label{subsec:non-commensurate}

The resulting effective model (\ref{eq:effective-fin}) simulated by the coupled Heisenberg one (\ref{eq:Heff}) gives rise to the spin squeezing generation  also for non-commensurate coupling phases~$\phi$, i.e. the one which is not equal to integer multiplications of $2 \pi/N$. In general, the results depend strongly on the chosen initial spin coherent state $|\theta, \varphi\rangle $ and parameters $\eta $ and $\gamma$.  

Let us discuss the situation when the initial spin coherent state  is polarized along the $z$ axis: $|0, 0\rangle = \bigotimes_{j=1}^N \ket{\uparrow}_j$.
Examples of the best squeezing and the best squeezing times are shown in Fig.~\ref{fig:fig6} (a)-(d) panels when $N=100$. A characteristic behavior is the OAT level of best squeezing for positive values of $\eta$ which is demonstrated in panel (b). In other cases, when $\eta$ is negative, the OAT level is also achieved mainly with $\eta$ close to zero, see e.g. in panels (c) and (d). 
It is possible to exceed the OAT level of squeezing when $\eta$ approaches the local minimum, see panels (a), (c), and (d). Interestingly, the last term in the effective model (\ref{eq:effective-fin}), namely $\gamma \hat{S}_x$, does not dominate the dynamics even if $\gamma$ is orders of magnitude larger than $\eta$. In Appendix~\ref{app:Incommensurate} we show the corresponding results for two different initial states. 
The OAT level of squeezing can be achieved when the initial state is polarized along the $y$-axis, $|\theta = \pi/2, \varphi =\pi/2\rangle$. The best squeezing and times are of the same level as the ones presented in Fig.~\ref{fig:fig6}. On the other hand, if the evolution starts with the state polarized along the $x$-axis, $|\theta = \pi/2, \varphi =0\rangle$, the dominant Zeeman-like term $\gamma \hat{S}_x$ in~(\ref{eq:effective-fin}) freezes the dynamics of the spin state and only weak spin squeezing is generated for non-commensurate phases.

\section{Conclusions and Summary}
\label{sec:conclusions}

We studied in detail the effect of OBC on the generation of spin squeezing in one-dimensional isotropic Heisenberg spin-1/2 chains induced by the position-dependent spin-flip coupling with off-set phase $\phi_0$~(\ref{eq:SFCj}). 
We extended the spin-wave theory for the case of OBC using the coordinate Bethe ansatz. 
We derived analytically the effective model in terms of the collective spin operators which describe the squeezing dynamics in the weak coupling regime. 
The resulting effective model obtained differs significantly from the one under PBC and, therefore, provides an example when the boundaries significantly modify the dynamics of the system.
To classify the squeezing scenarios, we distinguished two cases depending on the commensurability of the coupling phase~$\phi$ for well-defined off-set phase $\phi_0=\phi(N+1)/2$. When the coupling phase is commensurate, the dynamics of spin squeezing is well captured by the non-isotropic TACT if $\phi\ne\pi$ and OAT for $\phi= \pi$. The most general case of non-commensurate phase $\phi$ and arbitrary off-set phase $\phi_0$ still gives rise to the simulation of a squeezing model although not a conventional one. It is in contrary to the PBC case where the OAT model is simulated by the system independently of $\phi$.
Our analytical predictions were confirmed by the full many-body numerical simulations. 

The results presented here show how to produce entangled states in the isotropic spin-1/2 Heisenberg chains with nearest-neighbor interactions. This is possible by the addition of the position-dependent spin-flip coupling that is weak enough to maintain the dynamics within the Dicke manifold and strong enough to excite spin waves that are extended over the entire system, allowing "effective" all-to-all interaction between the individual spins. It is also worth adding that the dynamics of generated spin-squeezed states can be frozen at a desired time just by turning off the spin-flipping term. 
The results obtained can be verified experimentally by current state-of-the-art experiments with ultra-cold atoms.

\section*{ACKNOWLEDGMENTS}

We gratefully acknowledge discussions with B.~B.~Laburthe-Tolra, M. R. de Saint-Vincent and A. Sinatra. 
We thank O. Stachowiak for discussions and providing us Fig.~\ref{Afig:fig2}.
This work was supported by the European Social Fund (Project No. Nr 09.3.3-LMT-K-712-23-0035) under grant agreement with the Research Council of Lithuania (M.M.S.), 
the Polish National Science Centre project
DEC-2019/35/O/ST2/01873 (T.H.Y.)
and Grant No. 2019/32/Z/ST2/00016 through the project MAQS under QuantERA, which has received funding from the European Union’s Horizon
2020 research and innovation program under grant agreement no 731473 (E.W.).
M.~P. acknowledges the support of the Polish National Agency for Academic Exchange, the Bekker program no: PPN/BEK/2020/1/00317, and ERC AdG NOQIA; Ministerio de Ciencia y Innovation Agencia Estatal de Investigaciones (PGC2018-097027-B-I00/10.13039/501100011033, CEX2019-000910-S/10.13039/501100011033, Plan National FIDEUA PID2019-106901GB-I00, FPI, QUANTERA MAQS PCI2019-111828-2, QUANTERA DYNAMITE PCI2022-132919, Proyectos de I+D+I “Retos Colaboración” QUSPIN RTC2019-007196-7); MICIIN with funding from European Union NextGenerationEU(PRTR-C17.I1) and by Generalitat de Catalunya; Fundació Cellex; Fundació Mir-Puig; Generalitat de Catalunya (European Social Fund FEDER and CERCA program, AGAUR Grant No. 2021 SGR 01452, QuantumCAT \& U16-011424, co-funded by ERDF Operational Program of Catalonia 2014-2020); Barcelona Supercomputing Center MareNostrum (FI-2022-1-0042); EU Horizon 2020 FET-OPEN OPTOlogic (Grant No 899794); EU Horizon Europe Program (Grant Agreement 101080086 — NeQST), National Science Centre, Poland (Symfonia Grant No. 2016/20/W/ST4/00314); ICFO Internal “QuantumGaudi” project; European Union’s Horizon 2020 research and innovation program under the Marie-Skłodowska-Curie grant agreement No 101029393 (STREDCH) and No 847648 (“La Caixa” Junior Leaders fellowships ID100010434: LCF/BQ/PI19/11690013, LCF/BQ/PI20/11760031, LCF/BQ/PR20/11770012, LCF/BQ/PR21/11840013). Views and opinions expressed in this work are, however, those of the author(s) only and do not necessarily reflect those of the European Union, European Climate, Infrastructure and Environment Executive Agency (CINEA), nor any other granting authority. Neither the European Union nor any granting authority can be held responsible for them.

A part of the computations was carried out at the Centre of Informatics Tricity Academic Supercomputer \& Network.

\section*{Author contributions} 
THY performed check-in numerical simulations and provided all numerical results presented in the paper. MP performed preliminary many-body numerical  calculations. THY, EW and GJ provided the spin wave states for OBC presented in Appendix~\ref{app:sws}. THY, G\v{Z}, and EW contributed to a derivation of the effective Hamiltonian~(\ref{eq:effective-fin}). THY and G\v{Z} calculated analytically the off-set phase $\phi_0$ as discussed in Appendix~\ref{app:phi0}. G\v{Z}, DB, and GJ contributed to calculations of $\eta$ for commensurate phases as shown in Appendix~\ref{app:triple-cos-sum}. EW and GJ conceived the idea and guided the research. THY and EW wrote the first draft. All the authors contributed to the discussion of the results and the manuscript preparation and revision. 

\appendix

\section{Spin-waves states for OBC}\label{app:sws}

In this section, we are interested in spin-wave states which are eigenstates of the isotropic Heisenberg model, 
\begin{equation}\label{eqA:HSE}
\hat{H}_{\rm SE}=J_{\rm SE}\sum_{j=1}^{N-1}(S_{j}^{z}S_{j+1}^{z}+S_{j}^{y}S_{j+1}^{y}+S_{j}^{x}S_{j+1}^{x}-\frac{1}{4}),
\end{equation}
for $N$ spins and open boundary conditions.
In the following, we will show that the spin-wave states are given by Eq. (\ref{eq:swsobc}) of the main text, namely
\begin{equation}\label{eq:swsobc-pm}
|m, {q}\rangle = \pm \sqrt{N} c_{N/2, \pm m} \sum_{j=1}^N p^{(q)}_j \hat{S}^{\pm}_j | m\mp 1\rangle.
\end{equation}
In the above equation, the states $| m\mp 1\rangle$ are Dicke states while the usage of the on-site rising and lowering operators $\hat{S}_j^\pm$ corresponds to the two ways of definition of spin wave states. Note that $S_z |m, {q}\rangle = m |m, {q}\rangle $, as each term comprising the state-vector~(\ref{eq:swsobc-pm}) is characterized by the same spin projection $m$. Furthermore, $\hat S^2|m, {q}\rangle = S(S+1)|m, {q}\rangle $, with $S=N/2 - 1$. To see this we notice that the states~(\ref{eq:swsobc-pm}) are constructed in such a way that
\begin{equation}
\label{m,q-->q}
|m, {q}\rangle \propto \hat{S}_\pm^{N/2-1\pm m} |q\rangle^\pm ,
\end{equation}
where the state-vector $|q\rangle^\pm \equiv  | \mp (N/2-1),q\rangle $ corresponds to the minimum and maximum value of the spin projection $m = \mp (N/2-1)$. Since $[\hat S^2, \hat S_{\pm}]=0$, then 
\begin{equation}
\hat S^2|m, {q}\rangle \propto \hat{S}_\pm^{N/2-1\pm m} \hat S^2 |q\rangle^\pm ,
\end{equation}
Therefore, one needs to find the action of the operator $\hat S^2$ on the state-vector $| q\rangle^\pm$ which is
\begin{align}
	&\hat{S}^2 | q\rangle^\pm = \left(\hat{S}_z^2 + \hat{S}_z + \hat{S}_- \hat{S}_+\right)  | q\rangle^\pm \nonumber \\
	&= \left[ \left(\frac{N}{2}\right)^2 -  \frac{N}{2} \right] | q\rangle^\pm + \left(\sum_j p^{(q)}_j\right)  
	\hat{S}_{\pm } |N/2, \mp N/2\rangle.
\label{eq:lkastterm}
\end{align}
One can see that the state-vectors $|q\rangle^\pm $ are eigenstates of the $\hat{S}^2$ operator with the spin quantum number $S=N/2-1$ if the last term in (\ref{eq:lkastterm}) is zero, i.e.
\begin{equation}
\label{p_j-condition}
\sum_j p^{(q)}_j =0.    
\end{equation}
In that case the state-vectors $|m, {q}\rangle$ with an arbitrary $m$ are also the eigenstates of $\hat S^2$ with the quantum number  $S=N/2-1$. Note that the explicit form of the coefficients $p^{(q)}_j$ presented later in Eq.(\ref{eq:results}) do obey the condition \eqref{p_j-condition}.  

\begin{figure*}
	\begin{picture}(240, 100)
	\put(-110,105){(a)}
	\put(-140,0){\includegraphics[width=0.3\textwidth]{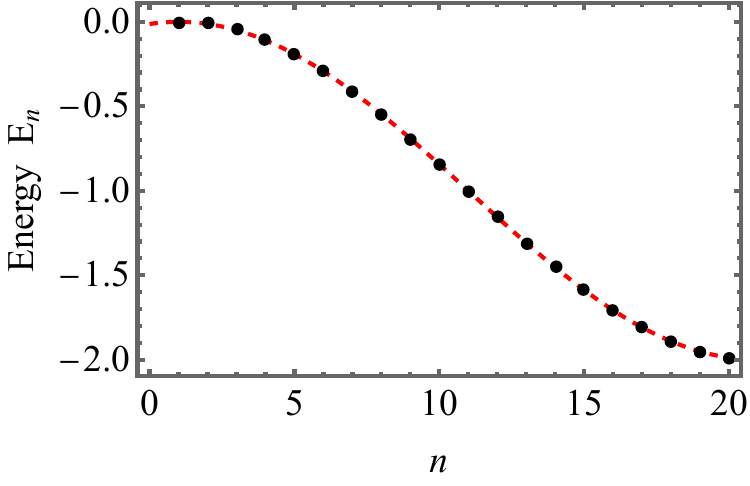}}
	\put(70,105){(b)}
	\put(40,0){\includegraphics[width=0.3\textwidth]{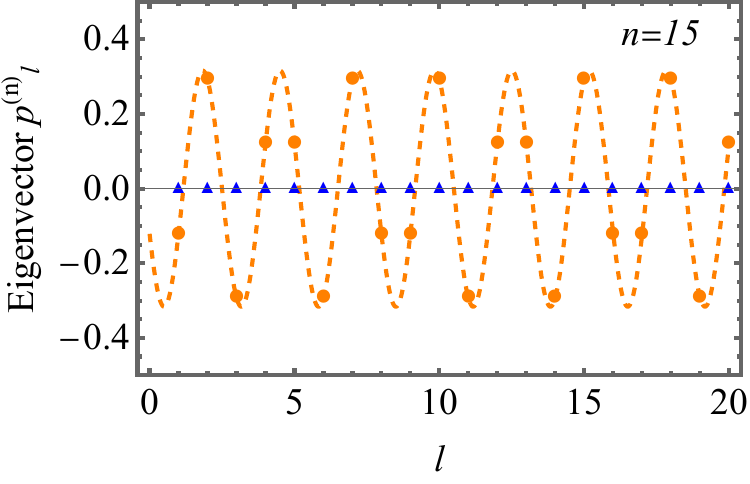}}
	\put(250,105){(c)}
	\put(220,0){\includegraphics[width=0.3\textwidth]{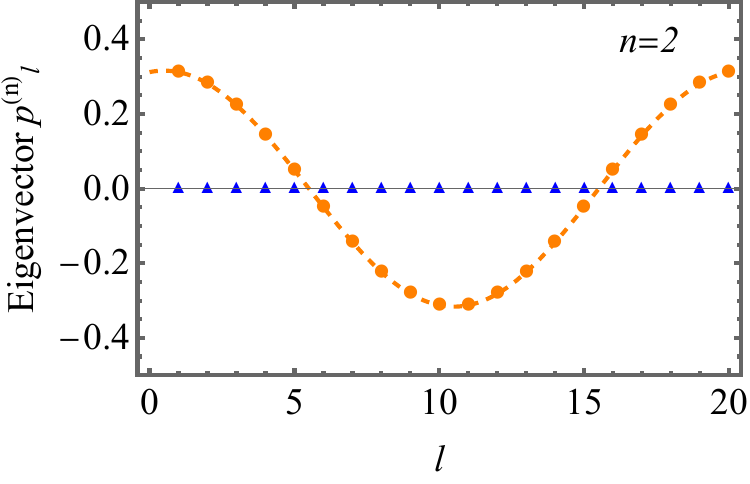}}
	\end{picture}
	\caption{(a) Energy spectrum $E_q$ of the spin-wave states for open boundary conditions, numerical (black points) and analytical (red dashed line) results. (b) and (c) show eigenvectors $p_l$  being solutions of (\ref{eq:A7})-(\ref{eq:A9}) for open boundary conditions when $q=15$ and $q=2$, respectively. Analytical results are marked by lines while the numerical one are marked by points (orange dashed lines mark real parts of $p_l$ while blue solid line are imaginary parts of $p_l$). An example for $N=20$.}
	\label{Afig:fig2}
\end{figure*}
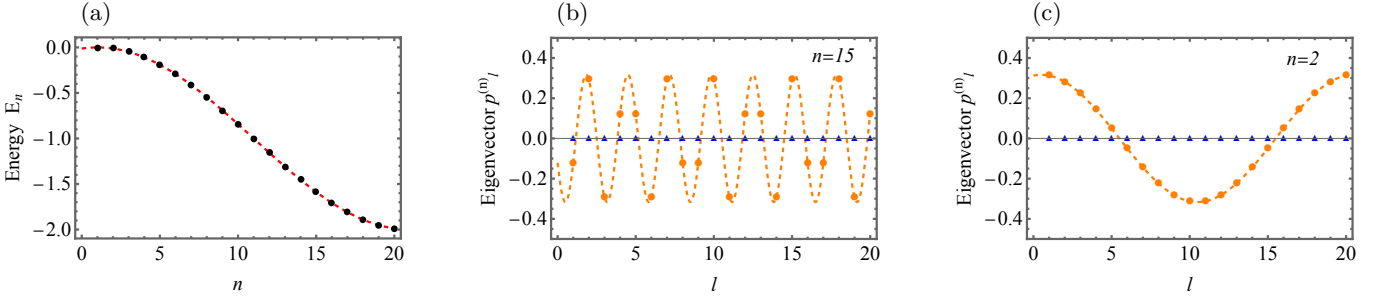

We are looking for the spin-wave states $|m, {q}\rangle$ which are eigenstates of the Hamiltonian~(\ref{eqA:HSE}). Since $[\hat{H}_{\rm SE}, \hat{S}_\pm] = 0$, using Eq.~\eqref{m,q-->q}, one can see that the eigenstates $|m, {q}\rangle$ of the Hamiltonian $\hat{H}_{\rm SE}$ have eigen-energies $E_q$ which do not depend on the quantum number $m$.
Therefore, by choosing the amplitudes $p_j^{(q)}$ in such a way that $|q\rangle^\pm $ are eigenstates of the spin exchange Hamiltonian (\ref{eqA:HSE}), the states $|m, {q}\rangle$ for any magnetization $m$ are also its eigenstates with the same eigen-energies $E_q$.

Below we show how to derive the form of $p_j^{(q)}$ for $|q\rangle^+ $ using OBC. The equations for $|q\rangle^- $ give the same expansion coefficients $p_j^{(q)}$ and the same eigen-energies~$E_q$.
Using the coordinate basis vectors: 
\begin{align}
|\tilde{l}\rangle &\equiv \hat{S}^+_l |-N/2\rangle=\hat{S}^+_l \bigotimes_{j=1}^N |\downarrow\rangle_j \,,
\end{align}
the spin wave states $|q\rangle^+$ can be represented as
\begin{align}
|q\rangle^+ &= \sum_{l=1}^N p_l |\tilde{l}\rangle.
\end{align}
The coefficients $p_l$ are evaluated by considering the eigenvalue problem 
\begin{equation}\label{eq:eigenvalueproblem}
(H - E I)\vec{p}=0, 
\end{equation}
where $I$ is the identity matrix, 
$\vec{p}=(p_1,p_2,...)$ and the matrix elements of $H$ are $H_{l', l} = \langle \tilde{l}'|\hat{H}_{SE}|\tilde{l}\rangle$.

The matrix form of eigenproblem (\ref{eq:eigenvalueproblem}) leads to the set of equations
\begin{align}
-\frac{J_{SE}}{2}p_1 + \frac{J_{SE}}{2}p_2 &= E p_1 \label{eq:A7}\\
\frac{J_{SE}}{2}p_{l-1} - J_{SE}p_{l} + \frac{J_{SE}}{2}p_{l+1} &= E p_l , \,\,\, {\rm for }\,\, l\in[2, N-1]\\
-\frac{J_{SE}}{2}p_{N} + \frac{J_{SE}}{2}p_{N-1} &= E p_N\label{eq:A9}
\end{align}
where (\ref{eq:A7}) and (\ref{eq:A9}) are for the boundary sites of the lattice. We use the idea by Puszkarski~\cite{PUSZKARSKI1973125} and add two virtual lattice sites $p_0$ and $p_{N+1}$ subject the boundary constrain $p_0 = p_1$ and $p_{N+1}=p_N$. In that case, the set of equations (\ref{eq:A7})-(\ref{eq:A9}) becomes equivalent to
the following set of bulk equations valid for any $l$:
\begin{equation}
\frac{J_{SE}}{2}p_{l-1} - J_{SE}p_{l} + \frac{J_{SE}}{2}p_{l+1} = E p_l .\label{eq:eq107}
\end{equation}
The solution to Eq.(\ref{eq:eq107}) can be represented as
\begin{equation}
\label{p_l}
p_l = p \cos\left[ k (l + u)\right],
\end{equation}
with the corresponding eigen-energies $E  = J_{SE} (\cos k - 1)$. The boundary constrain $p_0 = p_1$ requires $\cos(u k ) = \cos (u k + k)$ which is fulfilled for $u=-1/2$. The second constrain $p_{N+1}=p_L$ gives the requirement
\begin{equation}
\label{bc-right}
\cos(k N + k + uk)= \cos(k N + u k),
\end{equation}
which is fulfilled when $k=q \pi/N$, with $q=1,2,\cdots,N-1$ being an integer.
Therefore, we arrive at the required expansion coefficients and the corresponding eigen-energies: 
\begin{align}\label{eq:results}
p^{(q)}_l &= \sqrt{\frac{2}{N}} \cos\left[\frac{\pi}{N}\left( l -\frac{1}{2}\right)q \right], \\
E_{q} &= J_{SE} \left[\cos (\frac{\pi}{N} q) - 1\right].
\end{align}
Note, that the value $q=0$ is not included here, as in that case, the coefficients $p_l^{(q)}$ do not depend on $l$ and thus do not obey the condition (\ref{p_j-condition}). Although, such a state with $q=0$ is an eigenstate of the Hamiltonian $\hat{H}_{\rm SE}$, it belongs to the Dicke manifold and is characterized by the spin quantum number $S=N/2$ and zero eigen-energy.  

In Fig.~\ref{Afig:fig2} we show comparison of the numerical solution of (\ref{eq:A7})-(\ref{eq:A9}) with the analytical results. The perfect agreement can be noticed.

\section{Matrix representation of spin operators needed for effective model}\label{app:matrixspins}

In the following, we will present the matrix representation of various spin operators $\hat{S}_{\sigma}$ with $\sigma =z, \pm$, 
by using $\hat{S}_-|{S,m}\rangle=A_-^{S,m}|{S,m-1}\rangle$, $A_-^{S,m}=\sqrt{(S+m)(S-m+1)}$, $\hat{S}_+|{S,m}\rangle=A_+^{S,m}|{S,m+1}\rangle$, $A_+^{S,m}=\sqrt{(S-m)(S+m+1)}$.

The non-zero elements relevant for the relation of matrix representation with the corresponding spin operators, are
\begin{align}
&\langle N/2, m| \hat{S}_-^2| N/2, m+2\rangle =\nonumber \\
&\sqrt{(\frac{N}{2}+m+2)(\frac{N}{2}-m-1)(\frac{N}{2}+m+1)(\frac{N}{2}-m)}\label{eqA:B2}\\
&\langle N/2, m| \hat{S}_+^2| N/2, m-2\rangle =\nonumber \\
&\sqrt{(\frac{N}{2}+m)(\frac{N}{2}-m+1)(\frac{N}{2}+m-1)(\frac{N}{2}-m+2)} \label{eqA:B3}
\end{align}
One can show that the right hand site of Eq.(\ref{eqA:B2}) equals $(N-1) c_{N/2, m+1}^{-1} c_{N/2, -(m+1)}^{-1} $ and the right hand site of Eq.(\ref{eqA:B3}) equals $(N-1) c_{N/2, m-1}^{-1} c_{N/2, -(m-1)}^{-1} $. In addition, $\langle N/2, m|  \hat{S}_z^2| N/2, m\rangle = m^2 $
and $\langle N/2, m|  \hat{S}^2| N/2, m\rangle = \frac{N}{2} \left(\frac{N}{2} + 1\right)$ while 
$( c_{N/2, m}^{-2} + c_{N/2, -m}^{-2}) = \frac{2}{N-1} \left(m^2 + \frac{N}{2} + \frac{N^2}{4} \right)$.

\section{Effective model and off-set phase}\label{app:phi0}

The general form of the effective model including the first- and second-order perturbation terms is
\begin{align}
    \hat{H}_{\rm eff}&= 
    2 \chi_z \left(
    \hat{S}^2 + \hat{S}_z^2\right) 
    - {\rm Re} \left[\chi_x\right]
    \left( \hat{S}^2_+ + \hat{S}^2_-\right)\nonumber \\
    & - i {\rm Im} \left[\chi_x\right] \left( \hat{S}^2_+ - \hat{S}^2_-\right) +
v_x \hat{S}_x+
v_y \hat{S}_y
    \label{eq:effHopen}
\end{align}
which for $\phi_0=\phi (M+1)/2$ leads to (\ref{eq:effective-fin}).

While the general form of the effective Hamiltonian (\ref{eq:effHopen}) includes the mixed term $\hat{S}^2_+ - \hat{S}^2_-\propto \hat{S}_x\hat{S}_y + \hat{S}_y\hat{S}_x$ that complicates the effective model, it can be removed in general by a proper choice of the global phase factor in the atom-light coupling term. This is done by choosing a phase shift $\phi_0$ so that ${\rm Im} \left[\chi_x\right]  = 0$. In fact, it is sufficient to fulfill $\Im[(f_{q}^{\pm})^2] = 0$; $\forall q$ since $\Im[\chi_x] \propto \sum_{q} \left( \Im[(f_{q}^{\pm})^2] / E_{q} \right) $. 
By calculating explicitly 
\begin{equation}
	f_{q}^{\pm} = \sum_{j=1}^N p_j(q) \alpha^{\pm}_j = \frac{\sqrt{2}}{N}   \sum_{j=1}^N \cos\left[ \frac{\pi}{N}q\left( j-\frac{1}{2} \right) \right] e^{i\left( \phi j - \phi_0 \right)}, 
\end{equation}
using the geometric series result
\begin{equation}
	\sum_{j=1}^N r^j = \begin{cases}
		\frac{1-r^N}{r^{-1} - r} & \text{if } r \ne 1, \\
		N & \text{if } r = 1,
	\end{cases}
\end{equation}
we obtain
\begin{equation}
	f_{q}^{\pm} = \left\{
	\begin{array}{llr}
		\frac{e^{i\left( \frac{\phi}{2} -\phi_0  \right)}}{\sqrt{2}} \bigg[ 
		& \frac{e^{-i\pi\left(\frac{q}{2} - \frac{N\phi}{2\pi}\right)}}{N} 
		g(q, -\phi)
		&\\
		& +\frac{e^{i\pi\left(\frac{q}{2} + \frac{N\phi}{2\pi}\right)}}{N}
		g(q, \phi)
		\bigg]  & \text{if } \phi \ne \pm\frac{\pi}{N} q ,\\
		\frac{e^{i\left( \frac{\phi}{2} -\phi_0  \right)}}{\sqrt{2}}  && \text{if } \phi = \pm\frac{\pi}{N} q,
	\end{array}
	\right.
\end{equation}
where $g(q, \phi) = \frac{\sin\pi\left( \frac{q}{2} + \frac{N\phi}{2\pi} \right) }{\sin\frac{\pi}{N}\left( \frac{q}{2} + \frac{N\phi}{2\pi} \right) }$.  
This can also be written as
\begin{equation}
f_{q}^{\pm} = \left\{
	\begin{array}{llr}
 		\frac{e^{i\left( \frac{N+1}{2}\phi -\phi_0  \right)}}{\sqrt{2}} 
		\frac{i^{q}}{N} 
  &[ 
		(-1)^{q} g(q, -\phi)
  &\\
		&+ 
		g(q, \phi)
		],  & \text{if}\,\, \phi\ne \pm\frac{\pi}{N} q \\ 
	\frac{e^{i\left( \frac{\phi}{2} -\phi_0  \right)}}{\sqrt{2}}, && \text{if}\,\,
 \phi = \pm\frac{\pi}{N} q.  
\end{array}
	\right.
\end{equation}
Then
\begin{equation}
	\Im[\left(f_{q}^{\pm}  \right)^2 ] \propto \begin{cases}
	\sin \left( (N+1)\phi - 2\phi_0  \right)  & \text{if } \phi \ne \pm\frac{\pi}{N} q ,\\
	\sin (\phi -2\phi_0) & \text{if } \phi = \pm\frac{\pi}{N} q ,
	\end{cases}
\end{equation}
for $\Im[\left(f_{q}^{\pm}  \right)^2 ] = 0$; $\forall q$ it follows that
\begin{equation}
	\phi_0 = 
	\begin{cases}
		\frac{N+1}{2}\phi + \frac{\pi}{2} n & \text{if } \phi \ne \pm\frac{\pi}{N} q ,\\
		\frac{\phi}{2} + \frac{\pi}{2} n & \text{if } \phi = \pm\frac{\pi}{N} q,
	\end{cases}
\end{equation}
$\forall n \in \mathbb{Z}$. Notice we can write the second case result as the first one without any generality loss by changing the variable $n = q + n'$. As such, ${\rm Im}[\chi_{x}] = 0$ when
\begin{equation}
	\phi_0 = \frac{N+1}{2}\phi + \frac{\pi}{2}n; \quad \forall n \in \mathbb{Z}.
\end{equation}

\begin{figure*}[]
    \centering
    \includegraphics{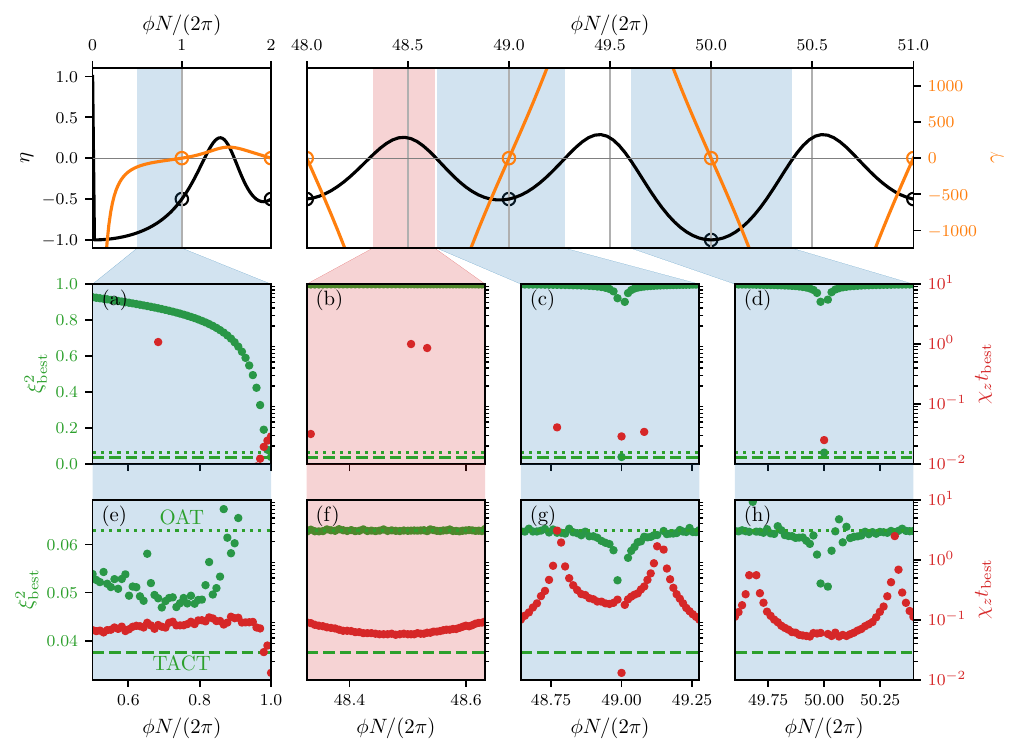}
    \caption{
    The best squeezing $\xi^2_\mathrm{best}$ (green points) and the best squeezing time $t_\mathrm{best}$ (red points) are shown in panels (a)-(d) for initial state $|\theta=\pi/2, \varphi=0\rangle$ and in panels (e)-(h) for initial state $|\theta=\pi/2, \varphi=\pi/2\rangle$. The numerical results for the effective model (\ref{eq:effective-fin}) with $N=100, J_\mathrm{SE} = 1$, $\Omega = |E_{q=1}|/10$, $\phi_0 = \phi(N+1)/2$ and $\eta > 0$ as indicated by the red shadowing areas and $\eta < 0$ indicated by the blue ones. The numerical values of $\eta$ and $\gamma$ used in simulations are shown in the top panels. The two limit cases for the values of $\xi^2_\mathrm{best}$, namely OAT and TACT for $N=100$, are marked with horizontal green dotted dashed lines, respectively.
   }
   \label{fig:best_sq_xy}
\end{figure*}

\section{Spin squeezing for incommensurate phase}
\label{app:Incommensurate}

We have showcased the best squeezing results for the initial coherent state $|\theta=0,\phi=0\rangle = \bigotimes_j \ket{\uparrow}_{j}$ in subsection~\ref{subsec:non-commensurate}, Fig.~\ref{fig:fig6}. Here we show that other choices for the initial state can provide different results. 
They are shown in Fig. \ref{fig:best_sq_xy} for
the initial state $|\theta=\pi/2, \varphi=0\rangle$ (middle panels) and 
$|\theta=\pi/2, \varphi=\pi/2\rangle$ (bottom panels).
The unitary evolution with the initial state being eigenstate of $\hat{S}_x$, it is $|\theta=\pi/2, \varphi=0\rangle$, shows practically no squeezing except very close to the commensurate phases or when $\gamma$ is very small, see in panels (a)-(d) Fig.~\ref{fig:best_sq_xy}.
On the other hand, when the initial state is eigenstate of 
$\hat{S}_y$, it is $|\theta=\pi/2, \varphi=\pi/2\rangle$, the squeezing dynamics is the same as for the initial state $|\theta=0,\phi=0\rangle$ which is presented in Fig.~\ref{fig:fig6}. This is shown in panels (e)-(h) of Fig.~\ref{fig:best_sq_xy}.

\section{Calculation of $\eta$ for commensurate phases}\label{app:triple-cos-sum}
For commensurate phase $\phi=2\pi n/N$, it is possible to calculate $\chi_z$ and $\chi_x$ analytically. Consequently, one can obtain $\eta$.

We make use of a method originally used in the study of random walks on lattices~\cite{Montrol1969, Lakatos1972} and also employed studying excitons in molecular aggregates~\cite{JCP1998JuzeliunasReineker}.

For convenience, let us represent Eqs.~\eqref{eq:Fdiag} and~\eqref{eq:FdiadFodd} in the following way:
\begin{equation}
\chi_z = \frac{\Omega^2}{4J_\mathrm{SE}(N-1)} F^{(\phi)}_{\rm diag},
\end{equation}
\begin{equation}
\chi_x = \frac{\Omega^2}{4J_\mathrm{SE}(N-1)} F^{(\phi)}_{\rm off},
\end{equation}
where we have defined the dimensionless sums $F^{(\phi)}_{\rm diag}$ and $F^{(\phi)}_{\rm off}$:
\begin{equation}
F_{\mathrm{diag}}^{(\phi)}=\frac{1}{N}\sum_{j,l=1}^{N}E_{j,l} \, e^{ \mathrm{i} \phi\left(j-l\right)},\label{eq:F_diag_Ejl}
\end{equation}
\begin{equation}
F_{\mathrm{off}}^{(\phi)} = \frac{1}{N}\sum_{j,l=1}^{N}E_{j,l}\, e^{ \mathrm{i} \phi\left(j+l\right)-\mathrm{i} 2\phi_{0}},\label{eq:F_off_Ejl}
\end{equation}
where
\begin{equation}
E_{j,l}=\frac{2}{N}\sum_{q=1}^{N}\frac{\cos\left[\frac{\pi q}{N}\left(j-\frac{1}{2}\right)\right] \cos\left[\frac{\pi q}{N}\left(l-\frac{1}{2}\right)\right]}{\cos(\pi q/N)-p}.
\end{equation}
Here we added the $q=N$ term which is zero, and introduced $p = 1 + \epsilon$ to avoid divergences. The limit $\epsilon \rightarrow 0^+$ will be taken at the end of calculations.

The main idea in finding this sum is to expand the denominator into a geometric series. To achieve this, one rewrites the denominator in the following way:
\begin{equation}
\cos(\pi q/N)-p = -\frac{b}{2}\left[1-b^{-1}e^{\mathrm{i}\pi q/N}\right]\left[1-b^{-1}e^{-\mathrm{i}\pi q/N}\right],
\end{equation}
where
\begin{equation}
b = p+\sqrt{p^{2}-1}.
\end{equation}
By using the symmetry of the summand to expand the summation limits, one can rewrite $E_{j,l}$ as
\begin{equation}
E_{j,l}=-C_{j+l-1}-C_{j-l}-\frac{1}{N}\frac{1}{1-p},\label{eq:E_jl_intermediate}
\end{equation}
where
\begin{equation}
C_{n}=\frac{1}{bN}\sum_{q=1-N}^{N}\frac{e^{\mathrm{i}\pi qn/N}}{\left[1-b^{-1}e^{\mathrm{i}\pi q/N}\right]\left[1-b^{-1}e^{-\mathrm{i}\pi q/N}\right]}
\end{equation}
with $C_{-n}=C_{n}^{*}$. Note that the last term in Eq.~\eqref{eq:E_jl_intermediate} cancels the added $q=0$ term in the summation.

Representing the denominator in terms of the geometric series,
one has:
\begin{equation}
C_{n}=\frac{1}{bN}\sum_{q=1-N}^{N}\sum_{r=0}^{\infty}\sum_{s=0}^{\infty}e^{\mathrm{i}\pi q\left(n+r-s\right)/N}b^{-\left(r+s\right)}.
\end{equation}
Using
\begin{equation}
\frac{1}{N}\sum_{q=1-N}^{N}e^{\mathrm{i}\pi q\left(n+r-s\right)/N}=2\sum_{m=-\infty}^{\infty}\delta_{n+r-s,2Nm},
\end{equation}
one obtains
\begin{align*}
C_{n}=\frac{2}{b}\sum_{m=-\infty}^{\infty}\sum_{r=0}^{\infty}\sum_{s=0}^{\infty}b^{-\left(r+s\right)}\delta_{n+r-s,2Nm}.
\end{align*}
Due to the Kronecker delta, the terms in the summation are non-zero only if $s=r+n-2Nm$ or equivalently if $r=s-n+2Nm$. Assuming that $0\le n<2N$, the integer $s=r+n-2Nm$ is $s\ge0$ if $m\le0$, whereas the integer $r=s-n+2Nm$ is $r\ge0$ if $m\ge1$. Therefore it is convenient to split the summation over $m$ into a part with $m<1$ and that with $m>0$, giving:
\begin{align}
C_{n}=\,&2b^{-1}\sum_{m=0}^{\infty}\sum_{r=0}^{\infty}b^{-\left(2r+2Nm+n\right)}+\\
&2b^{-1}\sum_{m=1}^{\infty}\sum_{s=0}^{\infty}b^{-\left(2s+2Nm-n\right)}.
\end{align}
After evaluating the geometric sums, one arrives at
\begin{equation}
C_{n}=\frac{2}{b-b^{-1}}\frac{b^{-\left|n\right|}+b^{-2N+\left|n\right|}}{1-b^{-2N}}
\end{equation}
where we have used the relation $C_{-n}=C_{n}^{*}$.

Taking the limit $\epsilon \rightarrow 0^+$, one obtains:
\begin{equation}
\begin{split}
-3NE_{j,l}=\,&1-3j+3j^{2}-3l+3l^{2}+3N-\\
&-6\max\left(j,l\right)N+2N^{2}.  
\end{split}    
\end{equation}
Therefore, one can rewrite Eq.~\eqref{eq:F_diag_Ejl} in terms of a double summation over $j>l$ and a single summation for $j=l$:
\begin{equation}
F_{\mathrm{diag}}^{(\phi)}=\frac{2}{N}\sum_{j=1}^{N}\sum_{l=1}^{j-1}E_{j,l}\, e^{\mathrm{i} \phi\left(j-l\right)} +\frac{1}{N}\sum_{j=1}^{N}E_{j,j}.
\end{equation}
Performing this summation, one obtains:
\begin{equation}
F_{\mathrm{diag}}^{(\phi)}=-\csc^{2}\left(\frac{\pi n}{N}\right).  
\end{equation}
Remembering that $\phi=2\pi n/N$, one can rewrite this into:
\begin{equation}
F_{\mathrm{diag}}^{(\phi)}=\frac{2}{\cos\phi-1},
\end{equation}
thus proving the identity mentioned in the main text.

As for $F_{\mathrm{off}}^{(\phi)}$, the steps are analogous,
first rewriting the sum~\eqref{eq:F_off_Ejl}:
\begin{equation}
\begin{split}
F_{\mathrm{off}}^{(\phi)}=\,&\frac{2}{N}\sum_{j=1}^{N}\sum_{l=1}^{j-1}E_{j,l}\, e^{ \mathrm{i}\phi\left(j+l\right)-\mathrm{i}2\phi_{0}} +\\
&\frac{1}{N}\sum_{j=1}^{N}E_{j,j}\, e^{ \mathrm{i} 2\phi j-\mathrm{i}2\phi_{0}}.
\end{split}
\end{equation}
For the initial phase $\phi_{0}=\phi\left(N+1\right)/2$, summation yields:
\begin{equation}
F_{\mathrm{off}}^{(\phi)}=\frac{1}{2}\csc^{2}\left(\frac{\pi n}{N}\right),
\end{equation}
or equivalently:
\begin{equation}
F_{\mathrm{off}}^{(\phi)}=-\frac{1}{\cos\phi-1},
\end{equation}
as expected.

Having both $F_{\mathrm{diag}}^{(\phi)}$ and $F_{\mathrm{off}}^{(\phi)}$, one can confirm that:
\begin{equation}
\eta = \frac{\mathrm{Re}\left[F_{\mathrm{off}}^{(\phi)}\right]}{F_{\mathrm{diag}}^{(\phi)}}=-\frac{1}{2},
\end{equation}
as clearly seen in Fig.~\ref{fig:fig3}.

The exceptional case of $\phi=\pi$ must be considered separately giving for $\phi_0 = \phi(N+1)/2$:
\begin{equation}
F^{(\pi)}_{\mathrm{diag}} = -F^{(\pi)}_{\mathrm{off}} = 1.
\end{equation}
In general, for any $\phi_{0}$ one has the following identities:
\begin{equation}
F_{\mathrm{off}}^{(\pi)}=\frac{1}{2}\, e^{i \left( \frac{2\pi n}{N}- 2\phi_{0}\right)}\csc^{2}\left(\frac{\pi n}{N}\right),
\end{equation}
or equivalently:
\begin{equation}
F_{\mathrm{off}}^{(\pi)}=-\frac{e^{i\left(\phi-2\phi_{0}\right)}}{\cos\phi-1}.
\end{equation}

\bibliography{bibliography}

\end{document}